\documentclass[11pt,a4paper]{article}
\usepackage{jcappub}
\pdfoutput=1
\usepackage{graphicx}
\usepackage{longtable}
\usepackage{color}

\title{Thermal Fluctuations of Dark Matter in Bouncing Cosmology}

\author{Changhong~Li}

\affiliation{Department of Astronomy, Key Laboratory of Astroparticle Physics of Yunnan Province, \\Yunnan University, No.2 Cuihu North Road, Kunming, China 650091}

\emailAdd{changhongli@ynu.edu.cn}

\abstract{We investigate the statistical nature of the dark matter particles produced in bouncing cosmology, especially, the evolution of its thermal  fluctuations. By explicitly deriving and solving the equation of motion of super-horizon mode, we fully determine the evolution of thermal perturbation of dark matter in a generic bouncing background. And we also show that the evolution of super-horizon modes is stable and will not ruin the background evolution of a generic bouncing universe till the Planck scale. Given no super-horizon thermal perturbation of dark matter appears in standard inflation scenario such as WIMP(-less) miracles, such super-horizon thermal perturbation of dark matter generated during the generic bouncing universe scenario may be significant for testing and distinguishing these two scenario in near future. }
\arxivnumber{0000.00000}

\begin{document}
\maketitle

\section{Introduction}
In history, a new theory of early universe used to be motivated by the initial condition problems of the last old one theory. For instance, to solve the initial condition problems of the Big Bang Theory, {\it i.e. Horizon} and {\it Flatness} problems, the inflation scenario with the idea modeling exponential expansion of universe using simple scalar fields are proposed~\cite{Guth:1980zm}. Remarkably, beyond this original motivation that solving the initial condition problems of Big Bang Theory, the inflation scenario has also led to a lot of great progress in the study of cosmology. Particularly, it generates a nearly scale invariant power curvature spectrum~\cite{Mukhanov:1990me} that agrees well with the current array of observations of Cosmic Microwave Background(CMB) spectra~\cite{Komatsu:2010fb, Ade:2013kta}. Coming along with such great success,  the inflation scenario also suffers from its own initial condition problem, the {\it initial singularity problem}~\cite{Borde:1993xh}(similar to the Big Bang Theory). It  motivates cosmologists to develop new theories of early universe to solve such initial condition problem. 

In recent years, to address this {\it initial singularity problem} of the inflation scenario, many efforts utilizing a generic bouncing feature of early universe have been made~\cite{Wands:1998yp, Khoury:2001wf, Steinhardt:2001st, Steinhardt:2002ih, Gasperini:2002bn, Creminelli:2006xe, Cai:2007qw, Cai:2008qw, Wands:2008tv, Bhattacharya:2013ut, Odintsov:2014gea}. According to these bouncing universe models, which used to be inspired by the underlying physics such as string theory and quantum loop theory, universe is bouncing from the contracting phase to expanding phase with non-zero size~\cite{Brandenberger:2016vhg, Novello:2008ra, Brandenberger:2012zb}.  Therefore, the initial singularity from which the inflation scenario inevitably suffers then can be avoided in the bouncing universe scenario.  Moreover,  a stable and scale-invariant curvature spectrum that is compatible with current observations and free of initial singularity and fine-tuning problems has been obtained recently in the bouncing-universe scenario~\cite{Li:2011nj, Li:2013bha}. 

Encouraging by this finding, {\it  Big Bounce Genesis}, the idea of using the mass and cross section of dark matter as a smoking gun signal for the existence of the bouncing universe, was first proposed in~\cite{Li:2014era}(also see~\cite{Cheung:2014nxi, Li:2014cba}). By explicitly computing the thermal equilibrium and out-of-equilibrium productions of dark matter, a novel relation of cross-section, mass and relic abundance of dark matter, $\Omega_\chi\propto \langle\sigma v\rangle_\chi m_\chi^2$, that compatible with the observational constraints of dark matter energy density fraction, are obtained in the generic bouncing universe scenario~\cite{Li:2014era}. This characteristic relation, which is different from the predictions, $\Omega_\chi\propto \langle\sigma v\rangle_\chi^{-1} m_\chi^0$, of the dark matter models in the inflation scenario such as WIMP and WIMP-less miracles~\cite{Scherrer:1985zt, Feng:2008ya, Kolb:1990vq,Gondolo:1990dk, Dodelson:2003ft}, can be used to check against the data from the worldwide effort of recent and near future (in-)direct and collider detections of dark matter~\cite{Bernabei:2008yi, Bernabei:2010mq, Bernabei:2013cfa, Ahmed:2008eu, Ahmed:2009zw,Ahmed:2010wy,Agnese:2013rvf, Akerib:2013tjd, Aprile:2012nq, Aprile:2013doa, Aalseth:2010vx, Xiao:2014xyn,Battiston:2014pqa, Aguilar:2014mma,2008Natur,Adriani:2013uda,Strong:2005zx}, to determine whether or not universe undergoes a big bounce at a very early stage of cosmic evolution~\cite{Cheung:2014pea}. Furthermore, the recently observed GeV Excess of Gamma ray signal from central Milk Way~\cite{Daylan:2014rsa} also can be explained and well fitted by using this characteristic relation to compute the annihilation process of dark matter in big bounce genesis~\cite{CFL_gevbu}.

In the view of fast development of the bouncing universe and  big bounce genesis, we are well motivated to investigate more details of the thermodynamic nature and perturbative properties of  dark matter in the bouncing universe scenario, which may impose further theoretical and observational constraints and provide new predictions on the big bounce genesis and bouncing universe scenario.

In this paper, we work out the whole evolution of the total energy and  thermal perturbations of dark matter in a generic bouncing universe. We prove that the evolution of super-horizon mode of dark matter thermal perturbation is stable and does not ruin out the cosmological background of a generic bouncing universe till Planck scale.  

Following are the contents and organization of this paper. In Section~\ref{sec:I}, we briefly review the Big Bounce Genesis, a generic scenario for dark matter production in bouncing universe, based on our previous works~\cite{Li:2014era,Cheung:2014nxi, Li:2014cba}. It is the fundament of the issues we investigate in this paper. In section~\ref{sec:II}, we calculate the total energy and total energy fluctuation of dark matter for a given sub-horizon volume in the bouncing cosmological background by utilizing conventional thermodynamics. The famous relation, $\langle\delta E_\chi\rangle^2/\langle E_\chi\rangle^2=N_\chi^{-1}$, is re-obtained. By applying these results into Big Bounce Genesis, the total energy of dark matter  is obtained as a function of the cross-section and mass of dark matter and the background temperature. And we show that, for the whole evolution of the bouncing universe, the total energy of dark matter  is sub-dominated and takes a nearly constant fraction among the total energy of cosmological background in the given sub-horizon volume.

In Section~\ref{sec:III}, we propose an integrated scheme to investigate the evolution of both super-horizon and sub-horizon thermal  perturbation modes in the contracting and expanding phase of a generic bouncing universe. At first, we compute the evolution of the sub-horizon thermal fluctuation mode of dark matter in the contracting phase by utilizing the $k-l$ correspondence under Fourier transformations.   Furthermore, we explicitly derive and solve the equation of motion for the super-horizon thermal perturbations of dark matter. Then, by utilizing the matching conditions on the horizon-crossing, the evolution of these super-horizon modes are fully determined with the knowledge of the sub-horizon modes. Afterwards, we use the matching conditions on the bouncing point to determine the evolution of super-horizon thermal fluctuation mode of dark matter in the expanding phase. Thus the whole evolution of the sub-horion and super-horizon thermal perturbations of dark matter are fully determined in both contracting and expanding phases of a generic bouncing universe.

In Section~\ref{sec:IV}, we prove that the evolution of  super-horizon modes is stable in a generic bouncing background till the Planck scale. The discussions, summary and prospects for the theoretical and observational issues of thermal perturbation of dark matter in a bouncing universe are made at the Section~\ref{sec:V}.  

\section{A brief review of Big Bounce Genesis}
\label{sec:I}
In big bounce genesis scenario~\cite{Li:2014era,Cheung:2014nxi, Li:2014cba}, dark matter particles are produced thermally by the annihilation of other light particles in the hot plasma of cosmological background. As long as universe contracts, the temperature of background is rising to be larger than the mass of dark matter particle, therefore, dark matter particles are produced efficiently during the contracting phase. According to the bouncing cosmology, at the end of the contracting phase, universe is bouncing to an expanding phase. In the early stage of expanding phase, the temperature of background is still larger than the mass of dark particle, so that the dark matter is also produced efficiently. As long as universe continue to expand, the temperature of the background, eventually, falls below the mass of dark matter, then the thermal production of dark matter is exponentially suppressed, and the freeze-out process of dark matter commences. Generically, such a thermal production mechanism is model independent and irrelevant to the details of realization of bounce~\cite{Cai:2011ci, Li:2014cba}. 

\begin{figure}[htp!]
\centering
\includegraphics[width=0.7\textwidth]{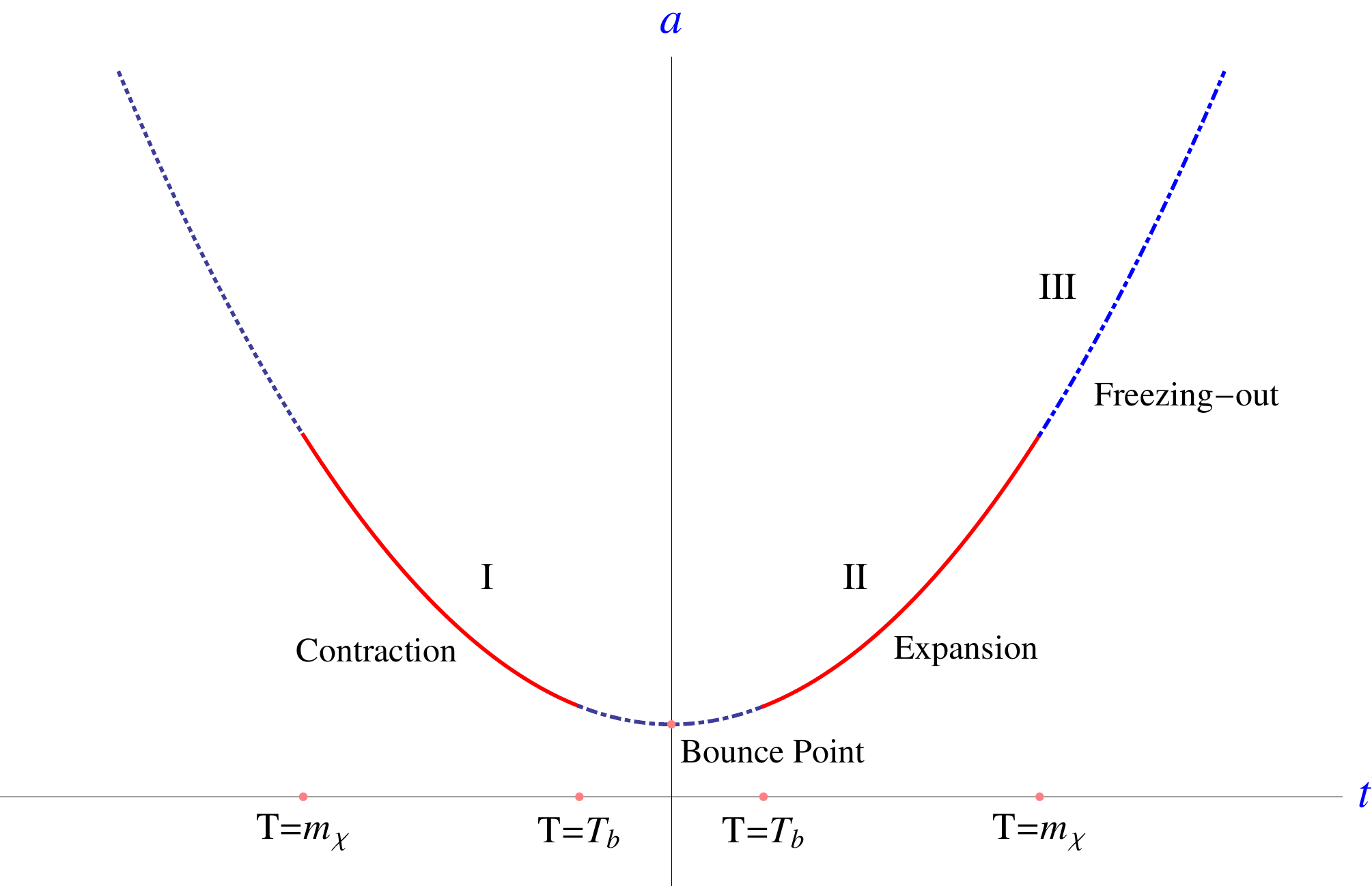}
\caption{Three stages for the dark matter production in a generic bouncing universe:  the pre-bounce contraction (phase I), the post-bounce expansion (phase II), and  the  freeze-out  of the dark matter particles (phase III), where the x-axis and y-axis are labeled by the physical time of cosmological evolution and the scale factor of universe respectively. 
}
\label{fig:cosbacofbus.pdf}
\end{figure}

For the thermal production of dark matter, a generic bouncing universe can be divided schematically into three stages~\cite{Li:2014era,Li:2014cba}, as following, as shown in Fig.~\ref{fig:cosbacofbus.pdf}.
%%%%%%%
\begin{enumerate}
\item{Phases I: the pre-bounce contraction, in which $H<0$ and $m_\chi\le T \le T_b$~;}
\item{Phases II:  the post-bounce expansion, in which $H>0$ and $m_\chi\le T \le  T_b$~;} 
\item{Phases III: the freeze-out phase of dark matter particles,  in which  $H>0$ and  $m_\chi \ge  T$~;}
\end{enumerate}
%%%%%%%
where $H$ is Hubble parameter taking negative value in the pre-bounce contraction and positive value in the post-bounce expansion and the freeze-out phase, $m_\chi$ the mass of dark matter. $T$ and $T_b$  are the temperatures of the cosmological background and of the short bounce process, respectively. In other words, $T_b$ is a critical temperature of the phase transition from standard physics regime into new physics one, which drives the bounce. Since the dark matter mass is assumed to be much lager than the energy scale of  the matter-radiation equality, these three stages are all radiation-dominated in which $H^2\propto a^{-4}\propto T^4$.

%%%%%%
%%%%%%
%%%%%%
%%%%%%

As shown in Fig.~\ref{fig:cosbacofbus.pdf}, the pre-bounce contraction and post-bounce expansion are connected by a short  bouncing process, which is highly model-dependent and we call it as bounce point for simplification. The details modeling of this bounce point, which mostly relies on more fundamental theories such as string phenomenology and quantum loop gravity, is an actively research field and have been hotly contested(for example see~\cite{Khoury:2001wf, Gasperini:2002bn, Creminelli:2006xe, Cai:2007qw, Wands:2008tv, Bars:2013vba, Bhattacharya:2013ut, Odintsov:2015ynk, Escofet:2015gpa, Haro:2015oqa,Gielen:2015uaa,Wan:2015hya, Oikonomou:2015qha, Nojiri:2015sfd, Cai:2015vzv, Qian:2016lbf, Odintsov:2016tar, Oikonomou:2016phk, Choudhury:2015baa, Cheung:2016oab, Nojiri:2016ygo, Wan:2015hya}). But luckily, for the purpose of analyzing the dark matter production, the effect around the bounce point is sub-leading since the cosmological evolution of the bounce point is assumed to be smooth and entropy-conserved~\cite{Cai:2011ci} and its characteristic time is very short comparing with the time-scale of dark matter production. Such consideration leads a matching condition for the number density of dark matter, $n_\chi$, at the bounce point,
\begin{equation}\label{eq:mcnddm}
n_\chi^-(T_b)=n_\chi^+(T_b).
\end{equation}
where superscripts $-$ and $+$ are used to label the contracting and expanding phase respectively. 
With this matching condition, the output of dark matter particles in the pre-bounce contraction becomes the initial abundance of dark matter in the post-bounce expansion. And before phase I in the bouncing universe scenario, the temperature of the background, $T\le m_\chi$ , is too low to produce dark-matter particles efficiently; the number density of dark-matter particles can thus be set to zero at the onset of the prebounce contraction 
phase,
\begin{equation}\label{eq:ivbb}
n_\chi^-(T=m_\chi)=0.
\end{equation}

By solving the Boltzman equation, which governs the production of dark matter particles,
\begin{equation} \label{eq:bmtno}
\frac{d(n_\chi a^3)}{a^3dt}=\langle\sigma v\rangle_\chi\left[\left(n_\chi^{(0)}\right)^2-n_\chi^2\right]~, 
\end{equation} 
with the initial condition Eq.(\ref{eq:ivbb}) and matching condition Eq.(\ref{eq:mcnddm}) , the details of evolution of number density of dark matter in a generic bouncing universe are unveiled in~\cite{Li:2014era}.
 
\begin{figure}[htp!]
\centering
\includegraphics[width=0.7\textwidth]{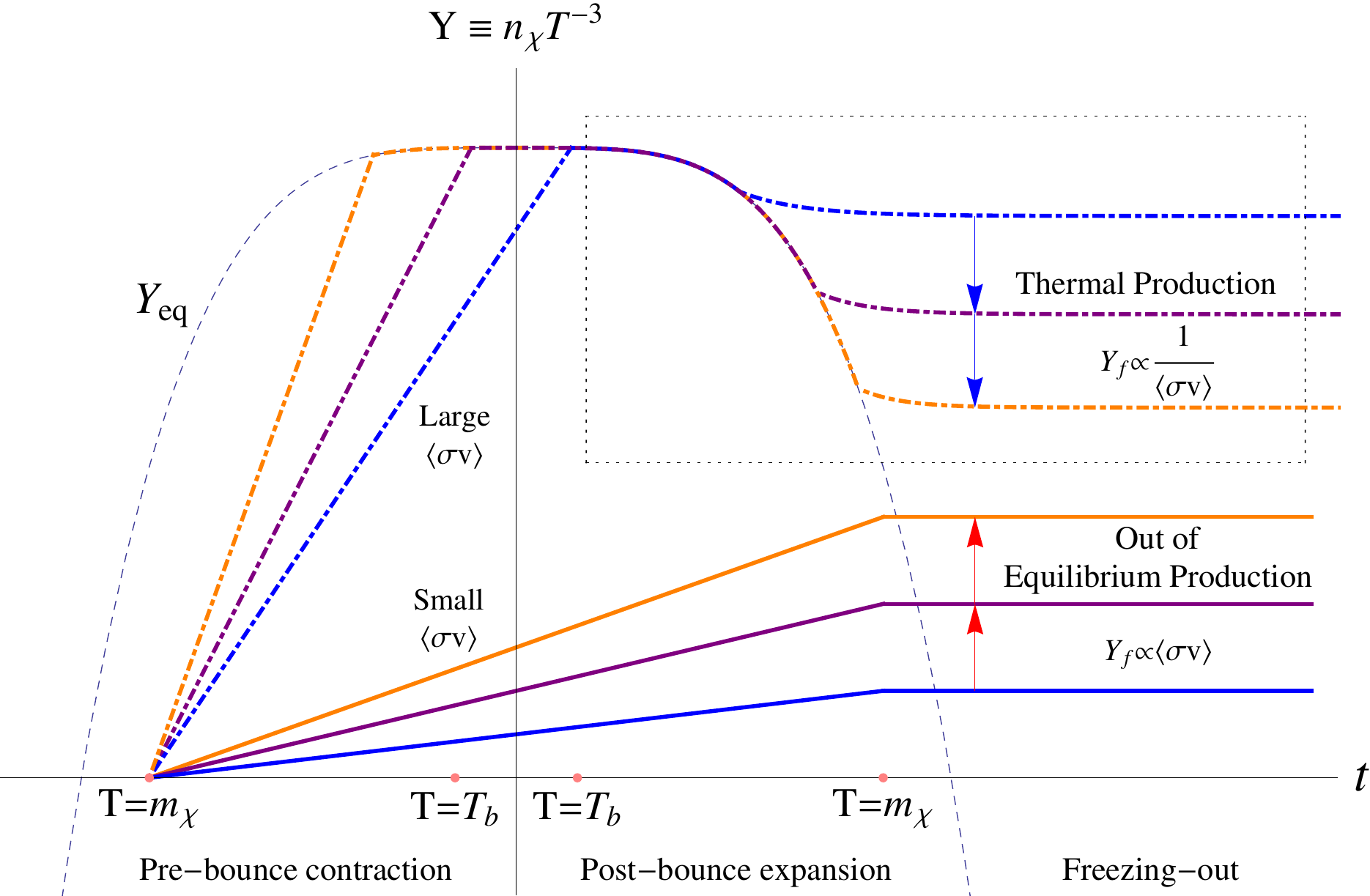}
\caption{The evolution of number density of the dark matter in a generic bouncing universe.}
\label{fig:Relice.pdf}
\end{figure}

It turns out that dark matter particles can be produced in two different avenues, thermal equilibrium production and out-equilibrium production, in a generic bouncing universe as shown in FIG.~\ref{fig:Relice.pdf}.
%%%%%%
%%%%%%
%%%%%%
%%%%%%

In the case of thermal equilibrium production(the upper dash-dotted lines in FIG.~\ref{fig:Relice.pdf}), the dark matter particle with large cross-section are produced very efficiently during the pre-bounce contracting phase and post-bounce expanding phase, so that the dark matter abundance achieves and tracks its thermal equilibrium value until the freezing-out phase.  Since the thermal equilibrium value of dark matter is fully determined by the parameters of cosmological background such as temperature, the relic abundance predicted by this case is identical to the prediction of the models in the standard inflationary cosmology such as WIMP and WIMP-less miracles(the dashed black frame in FIG.~\ref{fig:Relice.pdf})~\cite{Scherrer:1985zt, Feng:2008ya, Dodelson:2003ft, Li:2014era}. 

The novel case in big bounce genesis scenario is the out-equilibrium production avenue(the lower lines in FIG.~\ref{fig:Relice.pdf}). if the cross-section of dark matter particle is small enough, the production of dark matter is inefficient, so that its abundance can not achieve the thermal equilibrium value no matter how high the temperature of bounce is. In this case, dark matter takes a very weak freezing-out process that the informations of early universe evolution are preserved in the relic abundance of dark matter, which leads a new characteristic relation of dark matter cross-section and mass that compatible with current observations({\it c.f.} Branch B in Table II of~\cite{Li:2014era}),
\begin{equation}\label{eq:macsbbg}
\langle\sigma v\rangle=7.2\times 10^{-26}m_\chi^{-2}~,\quad m_\chi\gg 432~\text{eV}
\end{equation}  
where $\langle\sigma v\rangle$ is the thermally averaged cross-section of dark matter. As mentioned on above, this relation can be used to check against recent and near future data from experiments of dark matter to determine whether or not universe undergoes a big bounce at a very early stage of cosmic evolution~\cite{Cheung:2014pea,Vergados:2016niz}. And for more details of big bounce genesis, please refer the literatures~\cite{Li:2014era,Cheung:2014nxi, Li:2014cba}.

\section{Total Energy of Dark Matter in a Generic Bouncing Universe}
\label{sec:II}
{\bf Preliminary} Follow the conventional thermodynamics approach~\cite{Kadanoff:2000xz}, the canonical partition function of a single dark matter particle in relativistic limit, $p_\chi\gg m_\chi$, takes following form,
\begin{equation}\label{eq:zchi}
 z_\chi =(aL)^3\int\frac{d^3p_\chi}{(2\pi)^3}e^{-\beta \sqrt{p^2_\chi+m_\chi^2}}=\frac{C^3}{\pi^2}~,
\end{equation}
where $p_\chi$ and $m_\chi$ are the momentum and mass, respectively, of a single dark matter particle, $a$ is  the scale factor of universe, $L$ is a given  conformal length smaller than Hubble horizon $aL\le H^{-1}\equiv(\dot{a}/a)^{-1}$, $C\equiv\frac{aL}{\beta}$ is constant in an adiabatic cosmological background, and we take $\beta\equiv( k_BT)^{-1}$ and use natural units $h=c=k_B=1$ throughout this paper. Then the grand partition function $\mathcal{Z}$ can be written as,
\begin{equation}
\mathcal{Z_\chi}=\sum_{N=0}^{\infty}\frac{1}{N!}e^{\beta\mu N}z_\chi^N=\exp(e^{\beta\mu}z_\chi)
\end{equation}
where $\mu$ is the chemical potential, and $N!$ appears because the dark matter particles are considered to be indistinguishable. 
The total energy of dark matter is
\begin{equation} \label{eq:echi}
\langle E_\chi\rangle=-\frac{\partial}{\partial \beta}\ln \mathcal{Z}_\chi=-\mu e^{\beta\mu}z_\chi
\end{equation}
And the total energy fluctuation of dark matter in the given sub-horizon volume is
\begin{equation}\label{eq:dechi}
 \langle\delta E_\chi\rangle^2 =\langle E_\chi^2\rangle-\langle E_\chi\rangle^2=\frac{\partial^2\ln \mathcal{Z}_\chi}{\partial \beta^2}=\mu^2 e^{\beta\mu}z_\chi 
\end{equation}

{\bf Stability} By rewriting $\mu$ as a function of the equilibrium number density, $n_\chi^{(0)}=\int \frac{d^3k}{(2\pi)^3}e^{-\beta \sqrt{k^2_\chi+m_\chi^2}}$, and the number density, $n_\chi=e^{\beta\mu}n_\chi^{(0)}$, of dark matter~\cite{Dodelson:2003ft}, 
\begin{equation}\label{eq:mubeta}
\mu=\beta^{-1}\ln\left(\frac{n_\chi}{n_\chi^{(0)}}\right)~,
\end{equation}
the famous relation of $\langle\delta E_\chi\rangle^2/\langle E_\chi\rangle^2$ is re-derived,
\begin{equation}
\frac{\langle\delta E_\chi\rangle^2}{\langle E_\chi\rangle^2}=\frac{n_\chi^{(0)}}{n_\chi}z^{-1}_\chi=\frac{1}{N_\chi}~,
\end{equation}
where $N_\chi=n_\chi(aL)^3$ is the total number of dark matter particles and is much larger $1$ for the $L$ we interested in. It implies that the related thermodynamical quantities can be also well-defined in the bouncing universe scenario. And the total energy fluctuations of dark matter in a given sub-horizon volume, $aL\le H^{-1}$, is sub-dominated and stable, which will serves as the seeds for the super-horizon thermal perturbation we studied in next section.

{\bf Application} Now we can apply above results to Big Bounce Genesis\cite{Li:2014era}, in which the thermal equilibrium and out-of-equilibrium productions of dark matter have been studied in the generic bouncing universe scenario and  a novel relation of cross-section and mass of dark matter are predicted to satisfy the current astrophysical observations.  

According to Big Bounce Genesis\cite{Li:2014era}, in the radiation-dominated contracting and expanding phases of a generic bouncing universe, the number density of dark matter is
\begin{equation}\label{eq:rbi}
n_\chi=\frac{1-e^{-\lambda(1\mp y)}}{1+e^{-\lambda(1\mp y)}}n_\chi^{(0)},
\end{equation}
where $y\equiv m_\chi\beta$, $\lambda\equiv 2\pi^2f\langle\sigma v\rangle m_\chi $, $f\equiv\frac{m_\chi^2}{4\pi^2}(|H|y^2)^{-1}=1.5\times 10^{26}$~eV, and $\mp$ denotes the contracting and expanding phase respectively. $\bar{\langle \sigma v\rangle}$ and $\langle\sigma v\rangle$ are the thermally-averaged cross-section for dark matter particle, which is calculated in high temperature, $\beta\ll m_\chi$, and low temperature, $\beta\gg m_\chi$, limits respectively. For the dark matter produced through bosonic interaction, $\langle\sigma v\rangle$ is a constant and $\bar{\langle \sigma v\rangle} = \frac{1}{4}y^2\langle\sigma v\rangle$.

In the contracting phase($y\rightarrow 0$), the production processes of dark matter can be categorized into two different avenues, the {\it thermal equilibrium production} with $n_\chi\rightarrow n_\chi^{(0)}$ and the {\it out-equilibrium production} with $n_\chi\ll n_\chi^{(0)}$, according to Eq.(\ref{eq:rbi})~. Specifically, 
\begin{itemize}
\item {\it Thermal equilibrium production:}  For $\lambda \gg 1$, the production of dark matter is very efficient. With  Eq.(\ref{eq:rbi}),  the abundance of dark matter achieves its thermal equilibrium value, $n_\chi\rightarrow n_\chi^{(0)}$, during the contracting phase($y\rightarrow 0$);
\item {\it Out of equilibrium production:}  For $\lambda\ll 1$, the production of dark matter is  inefficient. Again, with Eq.(\ref{eq:rbi}), we can find that in this case the abundance of dark matter is much lower than its thermal equilibrium, $n_\chi\ll n_\chi^{(0)}$.
\end{itemize}
In~\cite{Li:2014era}, the dark matter candidates produced in both of these two avenues are proven to be able to satisfy the current observations of cosmological energy density fraction (see~\cite{Li:2014era, Cheung:2014nxi, Li:2014cba} for details). So we are studying the thermal fluctuation for both avenues in this paper, and from now on, we will use the terminology of the large $\lambda$($\lambda\gg 1$) and small $\lambda$ ($\lambda\ll 1$) to indicate the {\it thermal equilibrium production} and the {\it out-equilibrium production} respectively.

Substituting Eq.(\ref{eq:rbi}) into Eq.(\ref{eq:echi}),  the total energy of dark matter takes following form,
\begin{eqnarray}
 \langle E_\chi \rangle=\beta^{-1}\left(\frac{n_\chi}{n_\chi^{(0)}}\right)\left[-\ln\left(\frac{n_\chi}{n_\chi^{(0)}}\right)\right]z_\chi =\left\{  
\begin{array} {l}
 {\displaystyle \beta^{-1}\frac{\lambda(1\mp y)}{2} \ln\left[\frac{2}{\lambda(1\mp y)}\right]z_\chi,~\lambda\ll 1}   \\ 
 \\ 
  {\displaystyle \beta^{-1}2e^{-\lambda(1\mp y)}z_\chi, \qquad \qquad \qquad~\lambda\gg 1}    \\
\end{array}     
\right. .
\end{eqnarray}
For both of the large $\lambda$ and small $\lambda$ cases, their coefficients involving $\lambda$ are frozen in the high temperature contracting phase ($y \rightarrow 0$), and with  Eq.(\ref{eq:zchi}) we also recall that $z_\chi$ is constant, so that their total energy of dark matter $\langle E_\chi\rangle$ scales as $\beta^{-1}$ exactly. It indicates that the total energy of dark matter take a constant fraction among the total energy of the radiation-dominated background in which $E_\gamma$ also scales as $ \beta^{-1}$ exactly.

\section{Thermal Fluctuations of Dark Matter inside and beyond Hubble Horizon}
\label{sec:III}
Essentially, the energy density of the thermal fluctuations of dark matter can be determined by
\begin{equation}
\langle\delta\rho_\chi\rangle^2=\frac{\langle \delta E_\chi\rangle^2}{(aL)^6}~,
\end{equation}
and its power spectrum, $P(k)\equiv (2\pi^2)^{-1}k^3\delta\rho_k^2$, can be extracted from the $L$-dependence of $\delta\rho_\chi$ by utilizing Fourier transformation. 

However, this approach is only valid for the sub-horizon modes of the thermal fluctuations due to two reasons. (1) The thermodynamic computations of $\langle\delta E_\chi\rangle^2$ is done with a ``Hubble-adiabitiac" assumption that the wavelength of all modes is smaller than the length of the given volume, $a/k\le aL$, {\it i.e.} $k_{min}^{-1}=L$, and the value of $H$ is too small to affect the evolution of these short wavelength mode, $|H|<k_{min}/a=(aL)^{-1}\le k/a$. (2) Cosmologically, two points, which distance are larger than a Hubble horizon, are expected to be uncorrelated, so that the correlations of these original thermal fluctuations only exist insides the Hubble horizon, which also leads the requirement, $H\le k/a$. Even though these two reasons give out two almost same conditions, we should keep their difference in mind. The former reason concerns the modification of $k$-dependence of each thermodynamic quantity with a non-zero $H$, and the latter one tells us that in which space time volume, the original thermal fluctuations can emerge and correlate to each other. These two considerations are coincident in an expanding phase, $H>0$. But they are different in a contracting phase $H<0$ in which the second condition is always satisfied~\footnote{A similar discussion can be also used to explain why an effective horizon $|aH|^{-1}$ exits for the primordial perturbation in the contracting phase of a bouncing universe, even though the Hubble horizon $aH$ is obviously ``diverged''. }. And we call $(aH)^{-1}$ and $|aH|^{-1}$ as Hubble and effective horizon respectively when we need to distinguish them in some situations. To sum up, studying the evolution of the super-horizon ($|H|>k/a$) thermal perturbation modes requires a method beyond the conventional thermodynamics.      

In this paper, we propose a new scheme to investigate the evolution of both super-horizon and sub-horizon thermal  perturbation modes in the contracting and expanding phase of a generic bouncing universe. This scheme consists of the following four steps. 
\begin{itemize}
\item Step I: {\bf (Inside Horizon)} Computing the sub-horizon modes of thermal fluctuations of dark matter, $\delta\rho_k|_{k\ge |aH|}$, by utilizing the conventional thermodynamics and the $k-l$ correspondence, $k_{min}=L^{-1}$, under Fourier transformation;  

\item Step II: {\bf (Beyond Horizon)} Investigating the evolution of the super-horizon modes of thermal perturbations, $\delta\rho_k|_{k\le |aH|}$, by deriving and solving their equation of motion in long wavelength limit, and leaving the initial amplitude of these long wavelength perturbations undetermined; 

\item Step III: {\bf (Matching on Horizon Crossing)} As long as universe contracts, the effective horizon $|aH|^{-1}$ shrinks, so that the previously sub-horizon modes will becomes super-horizon after horizon crossing. Then we match the sub-horizon mode and the super-horizon mode on the moment of horizon crossing, $k=|aH|$, to determine the initial amplitude of the super-horizon thermal perturbations. Afterwards, the evolution of super-horizon thermal perturbation are fully determined during the contacting phase;

\item Step IV: {\bf (Matching on Bounce Point)} Eventually, universe are bouncing from the contracting phase to the expanding phase. By assuming the entropy of cosmological background are conserved before and after bouncing point, the matching conditions on the bouncing point are obtained. By utilizing these matching condition, the evolution of super-horizon thermal perturbation can be also fully determined during the expanding phase.

\end{itemize}

Following are the details of realization of these four steps.
%%%%%%%%%%%
%%%%%%%%%%%

\subsection{Inside Horizon} 

As discussed before, the sub-horizon mode of thermal fluctuation can be described completely by the conventional thermodynamics~\cite{Cai:2009rd, Biswas:2013lna}. The energy density of thermal fluctuation takes
\begin{equation}
\delta\rho_L^2=\frac{\langle\delta E_\chi\rangle^2}{(aL)^6}=\frac{\mu^2 e^{\beta\mu}}{\pi^{2}\beta^3}(aL)^{-3}~, \quad  aL\le |H|^{-1}~,
\end{equation}  
where $\delta\rho_L$ is short for $\langle\delta\rho_\chi\rangle_{aL}$ with subscript $aL$ denoting the physical length of the given volume. The $L$-dependence of $\delta\rho_L$ implies the distribution of amplitude for thermal fluctuation modes for each wavelength, which empowers us to go from the real space $L$ to the momentum space $k$,
\begin{equation}\label{eq:rlrk}
\delta \rho_L^2=\int_{k_{min}}^\infty\frac{k^2dk}{2\pi^2}\delta\rho_k^2~.
\end{equation}
By varying Eq.(\ref{eq:rlrk}) with respect to $l$ at $l=L$ and $k$ at $k=k_{min}$ repectively,
\begin{equation}
\left.\Delta(\delta \rho_l^2)\right|_{l=L}=\left.\frac{k^2\Delta k}{2\pi^2}\delta\rho_k^2\right|_{k=k_{min}}~,
\end{equation}
the power spectrum of the thermal fluctuation for the mode,  which physical wavelength is nearly equal to $aL$,   is then obtained
\begin{equation}
\delta\rho_{k_{min}}^2=\frac{6\pi^2\delta \rho_L^2}{k_{min}^3}=\frac{6\mu^2 e^{\beta\mu}}{(a\beta)^3}
\end{equation}
where we have used the relation $k_{min}=L^{-1}$~\footnote{A window function~(see~\cite{Biswas:2013lna} for the details) can be used to get the value of $\delta\rho_k$ for the modes which wave vector is much larger than $k_{min}$, $k\gg k_{min}$. And for the purpose of this paper, the approach taken at here is adequate and more straightforward.}. By taking $L$ go from $|aH|^{-1}$ to $0$, we obtain the power spectrum of the thermal fluctuations for all sub-horizon mode, 
\begin{equation}\label{eq:rksubh}
\delta\rho_{k}^2=\frac{6\pi^2\delta \rho_L^2}{k^3}=\frac{6\mu^2 e^{\beta\mu}}{(a\beta)^3}~,\quad k\ge |aH|
\end{equation}

Now we turn our attention to derive and solve the equation of motion for the long wavelength thermal perturbation mode. 
 
%%%%%%%%%%%
%%%%%%%%%%%
\subsection{Beyond Horizon} 
Much interesting issue of the thermal fluctuation of dark matter is the evolution of the super-horizon mode, which can be written in terms of the spatial variance of number density and average energy of the dark matter particle,  
\begin{equation}\label{eq:drh}
\delta\tilde{\rho}_\chi(x,t)=\delta n_\chi(x,t)\bar{\epsilon}_\chi(t)+n_\chi(t)\delta\bar{\epsilon}_\chi(x,t)
\end{equation}
where the $\tilde{}$ on $\delta\rho$ denotes the super-horizon mode, $\bar{\epsilon}_\chi\equiv \langle E_\chi\rangle/N_\chi$ is the average energy for one dark matter particle, and $\bar{\epsilon}_\chi=-\beta^{-1}\ln(n_\chi/n_\chi^{(0)})$ in the large temperature limit $y\rightarrow 0$. Being different from the sub-horizon thermal fluctuations originating from the thermal uncertainties and correlations in the grand ensemble, the super-horizon thermal perturbations, $\delta\tilde{\rho}_\chi(x,t)$ in Eq.(\ref{eq:drh}), describes how the energy density varies with the spatial variance of underlying physical quantities, such as local temperature and chemical potential.

Without loss of generality, we attributes all such thermal perturbation into the perturbation of temperature, 
\begin{equation}
\tilde{\beta}=\beta+\delta\beta(x,t)~.
\end{equation}
Then we have 
\begin{eqnarray}\label{eq:denchi}
&& \nonumber \delta n_\chi^{(0)}(x,t)=-3n_\chi^{(0)}\beta^{-1}\delta\beta(x,t)~,\\
&& \delta n_\chi (x,t)=n_\chi\left(\mu-3\beta^{-1}\right)\delta\beta(x,t)~,
\end{eqnarray}
and
\begin{equation} \label{eq:epmu}
\delta\bar{\epsilon}_\chi(x,t)=-\delta\mu(x,t)=0~.
\end{equation}
By substituting Eq.(\ref{eq:denchi}) and Eq.(\ref{eq:epmu}) into Eq.(\ref{eq:drh}), we obtain
\begin{equation}\label{eq:dtrrb}
\delta\tilde{\rho}_\chi(x,t)=-n_\chi\mu\left(\mu-3\beta^{-1}\right)\delta\beta(x,t)~.
\end{equation} 
Then by solving the equation of motion of $\delta \beta$, we can figure out the evolution of the super-horizon thermal perturbation of dark matter $\delta \tilde{\rho}_\chi$.

{\bf Equation of Motion}  Expanding the Boltzmann equation, which governs the evolution of dark matter in a generic bouncing universe,
\begin{equation} \label{eq:bmt}
\frac{d(\tilde{n}_\chi a^3)}{a^3dt}=\bar{\langle\sigma v\rangle}\left[\left(\tilde{n}_\chi^{(0)}\right)^2-\tilde{n}_\chi^2\right]~, 
\end{equation} 
up to the first order, we obtain the dynamic equation for the number density fluctuations,
\begin{equation} \label{eq:bmf}
\frac{d(\delta n_\chi a^3)}{a^3dt}=\bar{\langle\sigma v\rangle}\left[2 n_\chi^{(0)}\delta n_\chi^{(0)}-2n_\chi\delta n_\chi\right]~, 
\end{equation} 
where the $\tilde{A}$ includes the fluctuation and mean value of $A$, $\tilde{A}=A+\delta A(x,t)$, and $\delta A$ is short for $\delta A(x,t)$. Substituting Eq.(\ref{eq:denchi})  into Eq.(\ref{eq:bmf}), and using the zeroth order of Eq.(\ref{eq:bmt}), we obtain the equation of motion for the temperature fluctuation caused by the dark matter\footnote{In this paper, we focus on the thermal fluctuation of dark matter. The back-reaction from the scalar perturbation of metric are ignored for simplification.},  
\begin{eqnarray}\label{eq:pdebee}
&& \nonumber \bar{\langle\sigma v\rangle}\left[\left(n_\chi^{(0)}\right)^2-n_\chi^2\right]\left(\mu-3\beta^{-1}\right)\delta\beta+n_\chi\frac{d\left(\mu-3\beta^{-1}\right)}{dt}\delta\beta+n_\chi\left(\mu-3\beta^{-1}\right)\frac{d(\delta\beta)}{dt}\\
&& =-\bar{\langle\sigma v\rangle}\left[6\left(n_\chi^{(0)}\right)^2\beta^{-1}+2 n_\chi^2\left(\mu-3\beta^{-1}\right)\right]\delta\beta~.
\end{eqnarray}
By utilizing the relation $a\propto \beta$ in the radiation dominated phase to simplify Eq.(\ref{eq:pdebee}), the equation of motion for the thermal fluctuations is obtained, 
\begin{equation} \label{eq:ddbtdt}
\frac{\partial (\delta \beta)}{\partial t}+\frac{d x^j}{dt}\frac{\partial (\delta \beta)}{\partial x^j}+\Theta\delta \beta=0~.
\end{equation}
where $x^j$ are spatial coordinates with $j=1, 2, 3$, and $\Theta$ is defined as 
\begin{eqnarray}
 \Theta\equiv \left\{e^{-g(y)}\bar{\langle\sigma v\rangle}\frac{m_\chi^3}{\pi^2y^3}\left[1+e^{2g(y)} +6(g(y)-3)^{-1}\right]-[1-g^\prime y(g(y)-3)^{-1}]H\right\}
\end{eqnarray}
for short, $g(y)\equiv \ln\left(n_\chi/n_\chi^{(0)}\right)$, $g^\prime\equiv dg(y)/dy$,  and $H$ is the Hubble parameter, $H=\dot{a}/a$. 
  
{\bf Application} In the high temperature limit $y\ll 1$, $\Theta$ can be simplified to be, 
\begin{equation}\label{eq:stheq}
\Theta= \mathcal{C}y^{-1}-H~,
\end{equation}
where 
\begin{eqnarray}\label{eq:mceue}
&& \mathcal{C}=
\left\{  
\begin{array} {l}
 {\displaystyle   \qquad \frac{m_\chi^2}{4\pi^4}(1\mp y)^{-1}, \qquad \qquad \lambda\ll 1}   \\ 
 \\ 
  {\displaystyle  - \frac{(2+\pi^2)\langle\sigma v\rangle m_\chi^3}{3\pi^2}e^{-\lambda(1\mp y)}, ~\lambda\gg 1}    \\
\end{array}     
\right. 
\end{eqnarray}
where $\mp$ denotes the contracting and expanding phase respectively. 

Concerning with the developing of thermal fluctuation of dark matter during its production, we re-parameterize the physical time $t$ by $y$. And using the relation $a\propto \beta$ in the radiation dominated phase, we have 
 \begin{equation}\label{eq:dttody}
\frac{\partial (\delta\beta)}{\partial t}=Hy\frac{\partial (\delta \beta)}{\partial y}~.
\end{equation} 
By utilizing Eq.(\ref{eq:dttody}), Eq.(\ref{eq:ddbtdt}) can be re-written in respect to $y$,  
\begin{equation} \label{eq:ddbtdy}
\frac{\partial (\delta \beta)}{\partial y}+\frac{1}{yH}\frac{d x^j}{d t}\frac{\partial (\delta \beta)}{\partial x^j}+\frac{\Theta}{yH}\delta \beta=0 ~.
\end{equation} 
Furthermore, for the super-horizon mode of $\delta \beta$ which $k\le |aH|^{-1}$, $\delta \beta_{k}$, this equation can be simplified further in long wavelength limit,
\begin{equation}\label{eq:ddbdyl}
\frac{\partial (\delta \beta_{k})}{\partial y}+\frac{\Theta}{yH}\delta \beta_{k}=0 ~,
\end{equation}
with dropping out of the term $\partial (\delta \beta_{k})/\partial x^j$. 

Solving Eq.(\ref{eq:ddbdyl}) with Eq.(\ref{eq:stheq}), we get
\begin{equation}
 \delta \beta_{k}(t)= \delta \beta_{k}(t_i^\mp)\frac{y(t)}{y(t_i^\mp)}e^{Q(y)}
\end{equation}
where 
\begin{eqnarray}\label{eq:qyexp}
&& Q(y)=
\left\{
\begin{array}{l}
{\displaystyle  \frac{1}{\pi^2}\ln\left(\frac{1\mp y(t^\mp_i)}{1\mp y}\right),\quad\lambda \ll 1}\\
\\
{\displaystyle \frac{2(2+\pi^2)}{3\pi^2}\left[e^{-\lambda(1\mp y(t_i^\mp))}-e^{-\lambda(1\mp y )}\right],\quad \lambda\gg 1}\\
\end{array}
\right.
\end{eqnarray}
and we re-label each quantity $A$ by the physical time $t$ as $A(t)$, and $t_i$ is the moment of each super-horizon mode initiated on, and again the $\mp$ denotes the contracting and expanding phase respectively. 

In the high temperature limit, $ y \ll 1$, $e^{Q(y)}=1$ for $\lambda\gg1$ and $\lambda \ll 1$ in both of the contracting and expanding phases. Therefore, the perturbation of temperature evolves as following,
\begin{equation}\label{eq:betai}
\delta\beta_{k}(t)= \delta\beta_{k}(t_i^\mp)\frac{y(t)}{y(t_i^\mp)}~,
\end{equation}   
which is linear to $\beta$. Under Fourier transformations, Eq.(\ref{eq:dtrrb}) becomes
\begin{equation}\label{eq:dtrrbk}
\delta\tilde{\rho}_k(t)=-n_\chi\mu\left(\mu-3\beta^{-1}\right)\delta\beta_k(t)~.
\end{equation}
By substituting Eq.(\ref{eq:betai}) into Eq.(\ref{eq:dtrrbk}) and Eq.(\ref{eq:drh}), the energy density perturbation takes following form,
\begin{eqnarray}\label{eq:drhoi}
&& \nonumber \delta\tilde{\rho}_k(t)=-n_\chi\mu\left(\mu-3\beta^{-1}\right)\frac{y(t)}{y(t_i)}\delta\beta_{k}(t_i)\\
&&=\frac{n_\chi(t)\mu(t)\left(\mu(t)-3\beta(t)^{-1}\right)}{n_{\chi }(t_i^\mp)\mu(t_i^\mp)\left[\mu(t_i^\mp)-3\beta(t_i^\mp)^{-1}\right]}\frac{y(t)}{y(t_i^\mp)}\delta\tilde{\rho}_{k}(t_i^\mp).
\end{eqnarray} 
Therefore, if one can figure out the initial value of $\delta\tilde{\rho}_{k}$ and $\beta$, {\it i.e.} $\delta\tilde{\rho}_{k}(t_i^\mp)$ and $\beta(t_i^\mp)$, in the contracting and expanding phases respectively, by utilizing the matching conditions on the horizon crossing moment and bounce point, the evolution of the evolution of super-horizon thermal perturbation mode, $\delta\tilde{\rho}_{k}$, can be then fully determined.

Before turning our attention to study the matching condition on horizon crossing, we are simplifying Eq.(\ref{eq:drhoi}) in further by utilizing Eq.(\ref{eq:rbi}) that $n_\chi/n_\chi^{0}$ is frozen in the range of $y\ll 1$. Then Eq.(\ref{eq:drhoi}) becomes
\begin{equation}\label{eq:rkbfour}
 \delta\tilde{\rho}_k(t)=\left(\frac{\beta(t_i^\mp)}{\beta(t)}\right)^4\delta\tilde{\rho}_{k}(t_i^\mp)~.
\end{equation}
According to Eq.(\ref{eq:rkbfour}), during the contracting and expanding phase of the bouncing universe, each super-horizon mode scales as $\beta^{-4}$ and will not ruin the evolution of radiation-dominated background, which is also scaling as $\beta^{-4}$.

\subsection{Matching Conditions on Horizon Crossing}
During the contracting phase of universe, the effective horizon of universe shrinks, so that more and more sub-horizon modes of thermal fluctuations cross the horizon and become super-horizon thermal perturbation.  

The sub-horizon modes of thermal fluctuations with different value of $k$ will cross the effective horizon at different time. And the moment of each mode crossing the effective horizon is just the time of its corresponding super-horizon mode initiated, $t_i^-$, during the contracting phase. The horizon crossing condition is,
\begin{equation}\label{eq:hcatti}
k=|aH|_{t=t_i^-}~,
\end{equation} 
which leads $\beta(t_i)=\frac{\mathcal{C}_0}{4\pi^2f}k^{-1}$ with $\mathcal{C}_0\equiv\frac{a_i}{\beta_i}=0.752\times10^{-5}\text{eV}$.
And also the initial value of each super-horizon mode is determined by the value of sub-horizon mode at horizon crossing,
\begin{equation}\label{eq:rkti}
\delta\tilde{\rho}_{k}^2(t_i^-)=\left.\delta\rho_{k}^2\right|_{k= |aH|}=\left.\frac{6\mu^2 e^{\beta\mu}}{(a\beta)^3}\right|_{t=t_i^-}~,
\end{equation}   
where we have used Eq.(\ref{eq:rksubh}) with taking $k=|aH|$ at $t=t_i^-$~. 

{\bf Contracting Phase} Substituting the two matching conditions, Eq.(\ref{eq:hcatti}) and Eq.(\ref{eq:rkti}), on the horizon crossing into  Eq.(\ref{eq:rkbfour}), the evolution of super-horizon mode of thermal perturbation in the contracting phase is fully determined,
\begin{eqnarray}\label{eq:rkficp}
 \delta\tilde{\rho}_k(t)=\left(\frac{1}{\beta(t)}\right)^4 \sqrt{\frac{n_{\chi i}^-}{n_{\chi i}^{(0)-}}}\ln\left(\frac{n_{\chi i}^-}{n_{\chi i}^{(0)-}}\right)\frac{\sqrt{6}}{\mathcal{C}_0^\frac{3}{2}}=\left\{  
\begin{array} {l}
 {\displaystyle \left(\frac{1}{\beta(t)}\right)^4 \frac{\sqrt{6}}{\mathcal{C}_0^\frac{3}{2}}\frac{\lambda}{2} \ln\left(\frac{2}{\lambda}\right),~\lambda\ll 1}   \\ 
 \\ 
  {\displaystyle\left(\frac{1}{\beta(t)}\right)^4 \frac{\sqrt{6}}{\mathcal{C}_0^\frac{3}{2}} 2e^{-\lambda}, ~\qquad \lambda\gg 1}    \\
\end{array}     
\right. .
\end{eqnarray}
It turns out that the amplitude of thermal perturbation of dark matter are dependent on the value of $\lambda\equiv 2\pi^2f\langle\sigma v\rangle m_\chi$, {\it i.e.} the value of its cross-section and mass.
And the two avenues of dark matter production in bouncing cosmology, thermal equilibrium production and out-of-equilibrium production, can be distinguished in the level of thermal perturbation.

\subsection{Matching Conditions on Bouncing Point}

We have obtained the evolution of super-horizon mode of thermal perturbation in contracting phase, Eq.(\ref{eq:rkficp}), by utilizing the horizon crossing matching conditions, Eq.(\ref{eq:hcatti}) and Eq.(\ref{eq:rkti}).  To get the evolution of these super-horizon mode in the expanding phase, we need one more pair of matching conditions on the bouncing point. 
 
{\bf Expanding Phase}  Universe continues to be contracting until the bounce taking place. And after the bounce, universe evolves into the expanding phase~\footnote{The details of the bounce, which mostly relies on more fundamental theories such as string theory and quantum loop gravity, is model-dependent. However in general, the duration of the bounce is much shorter than the time-scale of dark matter production and the evolution of its thermal perturbations, {\it i.e.} $t_f^-\simeq t_b\simeq t_i^+ $, where $t_f^-$, $t_i^+$ and $t_b$ are, respectively, the moments of the contracting phase ending, the expanding phase beginning and the bounce taking place. Therefore, the effect on the evolution of thermal perturbation during the bounce is a sub-leading, which we have ignored for simplification.}. By assuming the entropy of the bounce are conserved~\cite{Cai:2011ci}, we have an additional pair of matching condition on the bounce,
\begin{equation}\label{eq:bomacob}
\beta(t_f^-)=\beta(t_i^+)
\end{equation}
and 
\begin{equation}\label{eq:bomacorho}
\delta\tilde{\rho}_k(t_f^-)=\delta\tilde{\rho}_k(t_i^+)
\end{equation}
where $t_f^-$ is the moment of the contracting phase ending.

Substituting Eq.(\ref{eq:bomacob}),  Eq.(\ref{eq:bomacorho}) and  Eq.(\ref{eq:rkficp})  into Eq.(\ref{eq:rkbfour}),  we obtain the evolution of super-horizon mode of thermal perturbation during the expanding phase,
\begin{eqnarray}\label{eq:rkfiep}
 \delta\tilde{\rho}_k(t)=\left(\frac{1}{\beta(t)}\right)^4 \sqrt{\frac{n_{\chi i}^-}{n_{\chi i}^{(0)-}}}\ln\left(\frac{n_{\chi i}^-}{n_{\chi i}^{(0)-}}\right)\frac{\sqrt{6}}{\mathcal{C}_0^\frac{3}{2}}~
=\left\{  
\begin{array} {l}
 {\displaystyle \left(\frac{1}{\beta(t)}\right)^4 \frac{\sqrt{6}}{\mathcal{C}_0^\frac{3}{2}}\frac{\lambda}{2} \ln\left(\frac{2}{\lambda}\right),~\lambda\ll 1}   \\ 
 \\ 
  {\displaystyle\left(\frac{1}{\beta(t)}\right)^4 \frac{\sqrt{6}}{\mathcal{C}_0^\frac{3}{2}} 2e^{-\lambda}, ~\qquad \lambda\gg 1}    \\
\end{array}     
\right. .
\end{eqnarray}
which have a same appearance of it in contracting phase.

\section{Stability of the  Super-horizon Thermal Fluctuation of Dark Matter}
\label{sec:IV}
According to Eq.(\ref{eq:rkbfour}), the energy density of each super-horizon mode increases at the same rate of the radiation dominated background, $\delta\tilde{\rho}_k\propto \rho_\gamma\propto \beta^{-4}$, where $\rho_\gamma$ is the energy density of cosmological background. It implies that, for each super-horzion mode, it does not ruin the evolution of background. However, as long as universe contracting, more and more thermal fluctuation modes become super-horzion. The accretion of super-horizon modes may substantially contribute to the energy density bills at the end of contracting phase $t=t_f^-=t_b$, 
\begin{eqnarray}\label{eq:rspint}
&& \nonumber \delta \tilde{\rho}^2(t_b)=\int_0^{k_b}\frac{k^2dk}{2\pi^2}\delta\tilde{\rho}_k^2(t_b)= \beta_b^{-11}\frac{n_{\chi i}^-}{n_{\chi i}^{(0)-}}\left[\ln\left(\frac{n_{\chi i}^-}{n_{\chi i}^{(0)-}}\right)\right]^2\frac{1}{64\pi^8f^3}~,
\end{eqnarray}
where we have used the relation $\beta_b=\frac{\mathcal{C}_0}{4\pi^2f}k_b^{-1}$ in last step.

The energy density of perturbation should be sub-dominated among the total energy density of cosmological background, 
\begin{equation}
\frac{\delta \tilde{\rho}^2(t_b)}{\rho_\gamma^2(t_b)}< 1
\end{equation}  
which leads a constraint on the highest temperature of bounce,
\begin{equation}
\frac{1}{(\beta_bM_p)^3}\frac{4f}{M_p}\frac{n_{\chi i}^-}{n_{\chi i}^{(0)-}}\left[\ln\left(\frac{n_{\chi i}^-}{n_{\chi i}^{(0)-}}\right)\right]^2<1~.
\end{equation}
Numerically,  $(n_{\chi i}^-/n_{\chi i}^{(0)-})\left[\ln(n_{\chi i}^-/n_{\chi i}^{(0)-})\right]^2\le\mathcal{O}(1)$ and $\frac{4f}{M_p}=0.14 \sim \mathcal{O}(1)$, so that this constraints can be satisfied until the background temperature approach to Planck scale,   
\begin{equation}
\beta_b M_p\le \mathcal{O}(1) \Longrightarrow T_b\le M_p~.
\end{equation}
It is proven that the evolution of super-horizon mode of dark matter thermal perturbation is stable and does not ruin out the cosmological background of a generic bouncing universe till Planck scale.

\section{Summary}
\label{sec:V}
In this paper, we present a detailed analysis of the thermodynamic nature and perturbative properties of dark matter in a generic bouncing universe scenario. 

Firstly, by applying the conventional thermodynamics to big bounce genesis, we work out the total energy and total energy fluctuations of dark matter, which is dependent on its cross-section, mass and background temperature, in a given sub-horizon volume of the bouncing cosmological background. We show that, for the whole evolution of the bouncing universe, the total energy of dark matter is sub-dominated and takes a nearly constant fraction among the total energy of cosmological background in the sub-horizon volume.

Moreover, for the super-horizon perturbation of dark matter that can not be described by the the conventional thermodynamics in Hubble adiabatic approximation, we propose an integrated scheme to investigate the evolution of both super-horizon and sub-horizon thermal  perturbation modes in the contracting and expanding phase of a generic bouncing universe. We explicitly derive and solve the equation of motion for the super-horizon thermal perturbations of dark matter. Then by utilizing the two group of matching conditions on the horizon-crossing and bouncing point respectively, the whole evolution of the super-horizon of dark matter thermal perturbations are fully determined in both contracting and expanding phases of a generic bouncing universe. And we prove that the evolution of super-horizon mode of dark matter thermal perturbation is stable and does not ruin out the cosmological background of a generic bouncing universe till Planck scale. 

To summarize, with the results presented in this paper, there are more works are worthy of being carried out in further for the issues of the thermodynamic nature and perturbative properties of dark matter in a generic bouncing universe scenario. We have determined the evolution of thermal perturbation of dark matter in the high temperature limit, $y\ll 1$, in which $Q(y)=1$ at Eq.(\ref{eq:qyexp}). But for the perturbations which exit effective horizon around $y\simeq 1$, $Q(y)$ is not generically equal to $1$. The numerical approach of this case may provide new signatures and physical implication for the the perturbations in big bounce genesis. Moreover, according to the results of this paper, Eq.(\ref{eq:rkficp}) and Eq.(\ref{eq:rkfiep}) , the two avenues of dark matter production in bouncing cosmology, thermal equilibrium production and  out-of-equilibrium production, can be distinguished in the level of thermal perturbation. Furthermore, such results of thermal perturbation also can be used to distinguish the thermal production avenue in bouncing cosmology and WIMP(-less) miracles in the inflation scenario even though they are degenerate in number density of dark matter. 

Furthermore, an analysis of the effects of such super-horizon thermal fluctuations of dark matter on the formation of large scale structure, clusters, galaxies and primordial blackhole is worthy to be investigated in near future~\cite{Navarro:1995iw, Zhao:2002rm,Diemand:2005vz,Frenk:2012ph,Bromm:2009uk, Umeda:2009yc, Colberg:2000zv,DiMatteo:2003zx, Maccio':2012uh}. And the back-reaction from the curvature perturbation, which may be subheading,  also should be taken account in further study.  Therefore, how to utilize these results to check against the data from astrophysical observations becomes a very exciting research issue.

\section{Acknowledgments}

We would like to thank Yeuk-Kwan E. Cheung, Yifu Cai and Vasilis Oikonomou for very useful discussions and comments. The work has been supported in parts by the National Natural Science Foundation of China (11603018), the Youth Grants of Fundamental Research of Yunnan Provincial Ministry of Science and Technology(2016FD006), the National Natural Science Foundation of China (11433004), the Leading Talents of Yunnan Province (2015HA022) and the Top Talents of Yunnan Province.

\clearpage
\addcontentsline{toc}{section}{References}

\bibliographystyle{JHEP}

\bibliography{Fluref}

\providecommand{\href}[2]{#2}\begingroup\raggedright\begin{thebibliography}{10}

\bibitem{Guth:1980zm}
A.~H. Guth, {\it {The Inflationary Universe: A Possible Solution to the Horizon
  and Flatness Problems}},  {\em Phys.Rev.} {\bf D23} (1981) 347--356.

\bibitem{Mukhanov:1990me}
V.~F. Mukhanov, H.~Feldman, and R.~H. Brandenberger, {\it {Theory of
  cosmological perturbations. Part 1. Classical perturbations. Part 2. Quantum
  theory of perturbations. Part 3. Extensions}},  {\em Phys.Rept.} {\bf 215}
  (1992) 203--333.

\bibitem{Komatsu:2010fb}
{\bf WMAP Collaboration} Collaboration, E.~Komatsu et~al., {\it {Seven-Year
  Wilkinson Microwave Anisotropy Probe (WMAP) Observations: Cosmological
  Interpretation}},  {\em Astrophys.J.Suppl.} {\bf 192} (2011) 18,
  [\href{http://xxx.lanl.gov/abs/1001.4538}{{\tt arXiv:1001.4538}}].

\bibitem{Ade:2013kta}
{\bf Planck} Collaboration, P.~Ade et~al., {\it {Planck 2013 results. XV. CMB
  power spectra and likelihood}},  {\em Astron.Astrophys.} {\bf 571} (2014)
  A15, [\href{http://xxx.lanl.gov/abs/1303.5075}{{\tt arXiv:1303.5075}}].

\bibitem{Borde:1993xh}
A.~Borde and A.~Vilenkin, {\it {Eternal inflation and the initial
  singularity}},  {\em Phys.Rev.Lett.} {\bf 72} (1994) 3305--3309,
  [\href{http://xxx.lanl.gov/abs/gr-qc/9312022}{{\tt gr-qc/9312022}}].

\bibitem{Wands:1998yp}
D.~Wands, {\it {Duality invariance of cosmological perturbation spectra}},
  {\em Phys.Rev.} {\bf D60} (1999) 023507,
  [\href{http://xxx.lanl.gov/abs/gr-qc/9809062}{{\tt gr-qc/9809062}}].

\bibitem{Khoury:2001wf}
J.~Khoury, B.~A. Ovrut, P.~J. Steinhardt, and N.~Turok, {\it {The Ekpyrotic
  universe: Colliding branes and the origin of the hot big bang}},  {\em
  Phys.Rev.} {\bf D64} (2001) 123522,
  [\href{http://xxx.lanl.gov/abs/hep-th/0103239}{{\tt hep-th/0103239}}].

\bibitem{Steinhardt:2001st}
P.~J. Steinhardt and N.~Turok, {\it {Cosmic evolution in a cyclic universe}},
  {\em Phys. Rev.} {\bf D65} (2002) 126003,
  [\href{http://xxx.lanl.gov/abs/hep-th/0111098}{{\tt hep-th/0111098}}].

\bibitem{Steinhardt:2002ih}
P.~J. Steinhardt and N.~Turok, {\it {A cyclic model of the universe}},  {\em
  Science} {\bf 296} (2002) 1436--1439.

\bibitem{Gasperini:2002bn}
M.~Gasperini and G.~Veneziano, {\it {The Pre - big bang scenario in string
  cosmology}},  {\em Phys.Rept.} {\bf 373} (2003) 1--212,
  [\href{http://xxx.lanl.gov/abs/hep-th/0207130}{{\tt hep-th/0207130}}].

\bibitem{Creminelli:2006xe}
P.~Creminelli, M.~A. Luty, A.~Nicolis, and L.~Senatore, {\it {Starting the
  Universe: Stable Violation of the Null Energy Condition and Non-standard
  Cosmologies}},  {\em JHEP} {\bf 0612} (2006) 080,
  [\href{http://xxx.lanl.gov/abs/hep-th/0606090}{{\tt hep-th/0606090}}].

\bibitem{Cai:2007qw}
Y.-F. Cai, T.~Qiu, Y.-S. Piao, M.~Li, and X.~Zhang, {\it {Bouncing universe
  with quintom matter}},  {\em JHEP} {\bf 0710} (2007) 071,
  [\href{http://xxx.lanl.gov/abs/0704.1090}{{\tt arXiv:0704.1090}}].

\bibitem{Cai:2008qw}
Y.-F. Cai, T.-t. Qiu, R.~Brandenberger, and X.-m. Zhang, {\it {A Nonsingular
  Cosmology with a Scale-Invariant Spectrum of Cosmological Perturbations from
  Lee-Wick Theory}},  {\em Phys.Rev.} {\bf D80} (2009) 023511,
  [\href{http://xxx.lanl.gov/abs/0810.4677}{{\tt arXiv:0810.4677}}].

\bibitem{Wands:2008tv}
D.~Wands, {\it {Cosmological perturbations through the big bang}},  {\em
  Adv.Sci.Lett.} {\bf 2} (2009) 194--204,
  [\href{http://xxx.lanl.gov/abs/0809.4556}{{\tt arXiv:0809.4556}}].

\bibitem{Bhattacharya:2013ut}
K.~Bhattacharya, Y.-F. Cai, and S.~Das, {\it {Lee-Wick radiation induced
  bouncing universe models}},  {\em Phys.Rev.} {\bf D87} (2013), no.~8 083511,
  [\href{http://xxx.lanl.gov/abs/1301.0661}{{\tt arXiv:1301.0661}}].

\bibitem{Odintsov:2014gea}
S.~D. Odintsov and V.~K. Oikonomou, {\it {Matter Bounce Loop Quantum Cosmology
  from $F(R)$ Gravity}},  {\em Phys. Rev.} {\bf D90} (2014), no.~12 124083,
  [\href{http://xxx.lanl.gov/abs/1410.8183}{{\tt arXiv:1410.8183}}].

\bibitem{Brandenberger:2016vhg}
R.~Brandenberger and P.~Peter, {\it {Bouncing Cosmologies: Progress and
  Problems}},  \href{http://xxx.lanl.gov/abs/1603.0583}{{\tt arXiv:1603.0583}}.

\bibitem{Novello:2008ra}
M.~Novello and S.~P. Bergliaffa, {\it {Bouncing Cosmologies}},  {\em
  Phys.Rept.} {\bf 463} (2008) 127--213,
  [\href{http://xxx.lanl.gov/abs/0802.1634}{{\tt arXiv:0802.1634}}].

\bibitem{Brandenberger:2012zb}
R.~H. Brandenberger, {\it {The Matter Bounce Alternative to Inflationary
  Cosmology}},  \href{http://xxx.lanl.gov/abs/1206.4196}{{\tt
  arXiv:1206.4196}}.

\bibitem{Li:2011nj}
C.~Li, L.~Wang, and Y.-K.~E. Cheung, {\it {Bound to bounce: a coupled
  scalar-tachyon model for a smooth cyclic universe}},  {\em Phys. of the Dark
  Universe} (2014) 18--33, [\href{http://xxx.lanl.gov/abs/1101.0202}{{\tt
  arXiv:1101.0202}}].

\bibitem{Li:2013bha}
C.~Li and Y.-K.~E. Cheung, {\it {The scale invariant power spectrum of the
  primordial curvature perturbations from the coupled scalar tachyon bounce
  cosmos}},  {\em JCAP} {\bf 1407} (2014) 008,
  [\href{http://xxx.lanl.gov/abs/1401.0094}{{\tt arXiv:1401.0094}}].

\bibitem{Li:2014era}
C.~Li, R.~H. Brandenberger, and Y.-K.~E. Cheung, {\it {Big Bounce Genesis}},
  {\em Phys.Rev.} {\bf D90} (2014), no.~12 123535,
  [\href{http://xxx.lanl.gov/abs/1403.5625}{{\tt arXiv:1403.5625}}].

\bibitem{Cheung:2014nxi}
Y.-K.~E. Cheung, J.~U. Kang, and C.~Li, {\it {Dark matter in a bouncing
  universe}},  {\em JCAP} {\bf 1411} (2014), no.~11 001,
  [\href{http://xxx.lanl.gov/abs/1408.4387}{{\tt arXiv:1408.4387}}].

\bibitem{Li:2014cba}
C.~Li, {\it {Thermally producing and weakly freezing out dark matter in a
  bouncing universe}},  {\em Phys. Rev.} {\bf D92} (2015), no.~6 063513,
  [\href{http://xxx.lanl.gov/abs/1404.4012}{{\tt arXiv:1404.4012}}].

\bibitem{Scherrer:1985zt}
R.~J. Scherrer and M.~S. Turner, {\it {On the Relic, Cosmic Abundance of Stable
  Weakly Interacting Massive Particles}},  {\em Phys.Rev.} {\bf D33} (1986)
  1585.

\bibitem{Feng:2008ya}
J.~L. Feng and J.~Kumar, {\it {The WIMPless Miracle: Dark-Matter Particles
  without Weak-Scale Masses or Weak Interactions}},  {\em Phys.Rev.Lett.} {\bf
  101} (2008) 231301, [\href{http://xxx.lanl.gov/abs/0803.4196}{{\tt
  arXiv:0803.4196}}].

\bibitem{Kolb:1990vq}
E.~W. Kolb and M.~S. Turner, {\it {The Early Universe}},  {\em Front.Phys.}
  {\bf 69} (1990) 1--547.

\bibitem{Gondolo:1990dk}
P.~Gondolo and G.~Gelmini, {\it {Cosmic abundances of stable particles:
  Improved analysis}},  {\em Nucl.Phys.} {\bf B360} (1991) 145--179.

\bibitem{Dodelson:2003ft}
S.~Dodelson, {\em {Modern cosmology}}.
\newblock Amsterdam, Netherlands: Academic Press, 2003.

\bibitem{Bernabei:2008yi}
{\bf DAMA} Collaboration, R.~Bernabei et~al., {\it {First results from
  DAMA/LIBRA and the combined results with DAMA/NaI}},  {\em Eur. Phys. J.}
  {\bf C56} (2008) 333--355, [\href{http://xxx.lanl.gov/abs/0804.2741}{{\tt
  arXiv:0804.2741}}].

\bibitem{Bernabei:2010mq}
{\bf DAMA, LIBRA} Collaboration, R.~Bernabei et~al., {\it {New results from
  DAMA/LIBRA}},  {\em Eur.Phys.J.} {\bf C67} (2010) 39--49,
  [\href{http://xxx.lanl.gov/abs/1002.1028}{{\tt arXiv:1002.1028}}].

\bibitem{Bernabei:2013cfa}
R.~Bernabei, P.~Belli, S.~d'Angelo, A.~Di~Marco, F.~Montecchia, et~al., {\it
  {Dark Matter investigation by DAMA at Gran Sasso}},  {\em Int.J.Mod.Phys.}
  {\bf A28} (2013) 1330022, [\href{http://xxx.lanl.gov/abs/1306.1411}{{\tt
  arXiv:1306.1411}}].

\bibitem{Ahmed:2008eu}
{\bf CDMS} Collaboration, Z.~Ahmed et~al., {\it {Search for Weakly Interacting
  Massive Particles with the First Five-Tower Data from the Cryogenic Dark
  Matter Search at the Soudan Underground Laboratory}},  {\em Phys. Rev. Lett.}
  {\bf 102} (2009) 011301, [\href{http://xxx.lanl.gov/abs/0802.3530}{{\tt
  arXiv:0802.3530}}].

\bibitem{Ahmed:2009zw}
{\bf CDMS-II} Collaboration, Z.~Ahmed et~al., {\it {Dark Matter Search Results
  from the CDMS II Experiment}},  {\em Science} {\bf 327} (2010) 1619--1621,
  [\href{http://xxx.lanl.gov/abs/0912.3592}{{\tt arXiv:0912.3592}}].

\bibitem{Ahmed:2010wy}
{\bf CDMS-II} Collaboration, Z.~Ahmed et~al., {\it {Results from a Low-Energy
  Analysis of the CDMS II Germanium Data}},  {\em Phys. Rev. Lett.} {\bf 106}
  (2011) 131302, [\href{http://xxx.lanl.gov/abs/1011.2482}{{\tt
  arXiv:1011.2482}}].

\bibitem{Agnese:2013rvf}
{\bf CDMS} Collaboration, R.~Agnese et~al., {\it {Silicon Detector Dark Matter
  Results from the Final Exposure of CDMS II}},  {\em Phys.Rev.Lett.} {\bf 111}
  (2013), no.~25 251301, [\href{http://xxx.lanl.gov/abs/1304.4279}{{\tt
  arXiv:1304.4279}}].

\bibitem{Akerib:2013tjd}
{\bf LUX Collaboration} Collaboration, D.~Akerib et~al., {\it {First results
  from the LUX dark matter experiment at the Sanford Underground Research
  Facility}},  {\em Phys.Rev.Lett.} {\bf 112} (2014) 091303,
  [\href{http://xxx.lanl.gov/abs/1310.8214}{{\tt arXiv:1310.8214}}].

\bibitem{Aprile:2012nq}
{\bf XENON100 Collaboration} Collaboration, E.~Aprile et~al., {\it {Dark Matter
  Results from 225 Live Days of XENON100 Data}},  {\em Phys.Rev.Lett.} {\bf
  109} (2012) 181301, [\href{http://xxx.lanl.gov/abs/1207.5988}{{\tt
  arXiv:1207.5988}}].

\bibitem{Aprile:2013doa}
{\bf XENON100} Collaboration, E.~Aprile et~al., {\it {Limits on spin-dependent
  WIMP-nucleon cross sections from 225 live days of XENON100 data}},  {\em
  Phys.Rev.Lett.} {\bf 111} (2013), no.~2 021301,
  [\href{http://xxx.lanl.gov/abs/1301.6620}{{\tt arXiv:1301.6620}}].

\bibitem{Aalseth:2010vx}
{\bf CoGeNT} Collaboration, C.~Aalseth et~al., {\it {Results from a Search for
  Light-Mass Dark Matter with a P-type Point Contact Germanium Detector}},
  {\em Phys.Rev.Lett.} {\bf 106} (2011) 131301,
  [\href{http://xxx.lanl.gov/abs/1002.4703}{{\tt arXiv:1002.4703}}].

\bibitem{Xiao:2014xyn}
{\bf PandaX} Collaboration, M.~Xiao et~al., {\it {First dark matter search
  results from the PandaX-I experiment}},  {\em Sci.China Phys.Mech.Astron.}
  {\bf 57} (2014) 2024--2030, [\href{http://xxx.lanl.gov/abs/1408.5114}{{\tt
  arXiv:1408.5114}}].

\bibitem{Battiston:2014pqa}
{\bf AMS} Collaboration, R.~Battiston, {\it {Precision measurements of $e^+
  e^?$ in Cosmic Ray with the Alpha Magnetic Spectrometer on the ISS}},  {\em
  Phys.Dark Univ.} {\bf 4} (2014) 6--9.

\bibitem{Aguilar:2014mma}
{\bf AMS} Collaboration, M.~Aguilar et~al., {\it {Electron and Positron Fluxes
  in Primary Cosmic Rays Measured with the Alpha Magnetic Spectrometer on the
  International Space Station}},  {\em Phys.Rev.Lett.} {\bf 113} (2014) 121102.

\bibitem{2008Natur}
J.~{Chang}, J.~H. {Adams}, H.~S. {Ahn}, G.~L. {Bashindzhagyan}, M.~{Christl},
  O.~{Ganel}, T.~G. {Guzik}, J.~{Isbert}, K.~C. {Kim}, E.~N. {Kuznetsov}, M.~I.
  {Panasyuk}, A.~D. {Panov}, W.~K.~H. {Schmidt}, E.~S. {Seo}, N.~V.
  {Sokolskaya}, J.~W. {Watts}, J.~P. {Wefel}, J.~{Wu}, and V.~I. {Zatsepin},
  {\it {An excess of cosmic ray electrons at energies of 300-800GeV}},  {\em
  Nature} {\bf 456} (Nov., 2008) 362--365.

\bibitem{Adriani:2013uda}
{\bf PAMELA} Collaboration, O.~Adriani et~al., {\it {Cosmic-Ray Positron Energy
  Spectrum Measured by PAMELA}},  {\em Phys.Rev.Lett.} {\bf 111} (2013) 081102,
  [\href{http://xxx.lanl.gov/abs/1308.0133}{{\tt arXiv:1308.0133}}].

\bibitem{Strong:2005zx}
A.~W. Strong, R.~Diehl, H.~Halloin, V.~Schoenfelder, L.~Bouchet, P.~Mandrou,
  F.~Lebrun, and R.~Terrier, {\it {Gamma-ray continuum emission from the inner
  galactic region as observed with integral/spi}},  {\em Astron. Astrophys.}
  {\bf 444} (2005) 495, [\href{http://xxx.lanl.gov/abs/astro-ph/0509290}{{\tt
  astro-ph/0509290}}].

\bibitem{Cheung:2014pea}
Y.-K.~E. Cheung and J.~Vergados, {\it {Direct dark matter searches - Test of
  the Big Bounce Cosmology}},  {\em JCAP} {\bf 1502} (2015), no.~02 014,
  [\href{http://xxx.lanl.gov/abs/1410.5710}{{\tt arXiv:1410.5710}}].

\bibitem{Daylan:2014rsa}
T.~Daylan, D.~P. Finkbeiner, D.~Hooper, T.~Linden, S.~K.~N. Portillo, N.~L.
  Rodd, and T.~R. Slatyer, {\it {The Characterization of the Gamma-Ray Signal
  from the Central Milky Way: A Compelling Case for Annihilating Dark Matter}},
   \href{http://xxx.lanl.gov/abs/1402.6703}{{\tt arXiv:1402.6703}}.

\bibitem{CFL_gevbu}
Y.-K.~E. Cheung, L.~Feng, and C.~Li, {\it {To appear}},  {\em To appear}
  (2016).

\bibitem{Cai:2011ci}
Y.-F. Cai, R.~Brandenberger, and X.~Zhang, {\it {Preheating a bouncing
  universe}},  {\em Phys.Lett.} {\bf B703} (2011) 25--33,
  [\href{http://xxx.lanl.gov/abs/1105.4286}{{\tt arXiv:1105.4286}}].

\bibitem{Bars:2013vba}
I.~Bars, P.~J. Steinhardt, and N.~Turok, {\it {Cyclic Cosmology, Conformal
  Symmetry and the Metastability of the Higgs}},  {\em Phys. Lett.} {\bf B726}
  (2013) 50--55, [\href{http://xxx.lanl.gov/abs/1307.8106}{{\tt
  arXiv:1307.8106}}].

\bibitem{Odintsov:2015ynk}
S.~D. Odintsov and V.~K. Oikonomou, {\it {Big-Bounce with Finite-time
  Singularity: The $F(R)$ Gravity Description}},
  \href{http://xxx.lanl.gov/abs/1512.0478}{{\tt arXiv:1512.0478}}.

\bibitem{Escofet:2015gpa}
A.~Escofet and E.~Elizalde, {\it {Gauss-Bonnet Modified Gravity Models with
  Bouncing Behavior}},  \href{http://xxx.lanl.gov/abs/1510.0584}{{\tt
  arXiv:1510.0584}}.

\bibitem{Haro:2015oqa}
J.~Haro, A.~N. Makarenko, A.~N. Myagky, S.~D. Odintsov, and V.~K. Oikonomou,
  {\it {Bouncing loop quantum cosmology in Gauss-Bonnet gravity}},  {\em Phys.
  Rev.} {\bf D92} (2015), no.~12 124026,
  [\href{http://xxx.lanl.gov/abs/1506.0827}{{\tt arXiv:1506.0827}}].

\bibitem{Gielen:2015uaa}
S.~Gielen and N.~Turok, {\it {A Perfect Bounce}},
  \href{http://xxx.lanl.gov/abs/1510.0069}{{\tt arXiv:1510.0069}}.

\bibitem{Wan:2015hya}
Y.~Wan, T.~Qiu, F.~P. Huang, Y.-F. Cai, H.~Li, and X.~Zhang, {\it {Bounce
  Inflation Cosmology with Standard Model Higgs Boson}},  {\em JCAP} {\bf 1512}
  (2015), no.~12 019, [\href{http://xxx.lanl.gov/abs/1509.0877}{{\tt
  arXiv:1509.0877}}].

\bibitem{Oikonomou:2015qha}
V.~K. Oikonomou, {\it {Singular Bouncing Cosmology from Gauss-Bonnet Modified
  Gravity}},  {\em Phys. Rev.} {\bf D92} (2015), no.~12 124027,
  [\href{http://xxx.lanl.gov/abs/1509.0582}{{\tt arXiv:1509.0582}}].

\bibitem{Nojiri:2015sfd}
S.~Nojiri, S.~D. Odintsov, and V.~K. Oikonomou, {\it {Unimodular $F(R)$
  Gravity}},  {\em JCAP} {\bf 1605} (2016), no.~05 046,
  [\href{http://xxx.lanl.gov/abs/1512.0722}{{\tt arXiv:1512.0722}}].

\bibitem{Cai:2015vzv}
Y.-F. Cai, F.~Duplessis, D.~A. Easson, and D.-G. Wang, {\it {Searching for a
  matter bounce cosmology with low redshift observations}},  {\em Phys. Rev.}
  {\bf D93} (2016), no.~4 043546,
  [\href{http://xxx.lanl.gov/abs/1512.0897}{{\tt arXiv:1512.0897}}].

\bibitem{Qian:2016lbf}
P.~Qian, Y.-F. Cai, D.~A. Easson, and Z.-K. Guo, {\it {Magnetogenesis in
  bouncing cosmology}},  \href{http://xxx.lanl.gov/abs/1607.0657}{{\tt
  arXiv:1607.0657}}.

\bibitem{Odintsov:2016tar}
S.~D. Odintsov and V.~K. Oikonomou, {\it {Deformed Matter Bounce with Dark
  Energy Epoch}},  \href{http://xxx.lanl.gov/abs/1606.0368}{{\tt
  arXiv:1606.0368}}.

\bibitem{Oikonomou:2016phk}
V.~K. Oikonomou, {\it {Reconstructing the evolution of the Universe from loop
  quantum cosmology scalar fields}},  {\em Phys. Rev.} {\bf D94} (2016), no.~4
  044004, [\href{http://xxx.lanl.gov/abs/1607.0710}{{\tt arXiv:1607.0710}}].

\bibitem{Choudhury:2015baa}
S.~Choudhury and S.~Banerjee, {\it {Hysteresis in the Sky}},  {\em Astropart.
  Phys.} {\bf 80} (2016) 34--89, [\href{http://xxx.lanl.gov/abs/1506.0226}{{\tt
  arXiv:1506.0226}}].

\bibitem{Cheung:2016oab}
Y.-K.~E. Cheung, X.~Song, S.~Li, Y.~Li, and Y.~Zhu, {\it {A smoothly bouncing
  universe from String Theory}},  \href{http://xxx.lanl.gov/abs/1601.0380}{{\tt
  arXiv:1601.0380}}.

\bibitem{Nojiri:2016ygo}
S.~Nojiri, S.~D. Odintsov, and V.~K. Oikonomou, {\it {Bounce universe history
  from unimodular $F(R)$ gravity}},  {\em Phys. Rev.} {\bf D93} (2016), no.~8
  084050, [\href{http://xxx.lanl.gov/abs/1601.0411}{{\tt arXiv:1601.0411}}].

\bibitem{Vergados:2016niz}
J.~D. Vergados, C.~C. Moustakidis, Y.-K.~E. Cheung, H.~Ejri, Y.~Kim, and
  Y.~Lie, {\it {Light WIMP searches involving electron scattering}},
  \href{http://xxx.lanl.gov/abs/1605.0541}{{\tt arXiv:1605.0541}}.

\bibitem{Kadanoff:2000xz}
L.~P. Kadanoff, {\em {Statistical physics: Statics, dynamics and
  renormalization}}.
\newblock 2000.

\bibitem{Cai:2009rd}
Y.-F. Cai, W.~Xue, R.~Brandenberger, and X.-m. Zhang, {\it {Thermal
  Fluctuations and Bouncing Cosmologies}},  {\em JCAP} {\bf 0906} (2009) 037,
  [\href{http://xxx.lanl.gov/abs/0903.4938}{{\tt arXiv:0903.4938}}].

\bibitem{Biswas:2013lna}
T.~Biswas, R.~Brandenberger, T.~Koivisto, and A.~Mazumdar, {\it {Cosmological
  perturbations from statistical thermal fluctuations}},  {\em Phys. Rev.} {\bf
  D88} (2013), no.~2 023517, [\href{http://xxx.lanl.gov/abs/1302.6463}{{\tt
  arXiv:1302.6463}}].

\bibitem{Navarro:1995iw}
J.~F. Navarro, C.~S. Frenk, and S.~D.~M. White, {\it {The Structure of cold
  dark matter halos}},  {\em Astrophys. J.} {\bf 462} (1996) 563--575,
  [\href{http://xxx.lanl.gov/abs/astro-ph/9508025}{{\tt astro-ph/9508025}}].

\bibitem{Zhao:2002rm}
D.~Zhao, H.~Mo, Y.~Jing, and G.~Boerner, {\it {The growth and structure of dark
  matter haloes}},  {\em Mon. Not. Roy. Astron. Soc.} {\bf 339} (2003) 12--24,
  [\href{http://xxx.lanl.gov/abs/astro-ph/0204108}{{\tt astro-ph/0204108}}].

\bibitem{Diemand:2005vz}
J.~Diemand, B.~Moore, and J.~Stadel, {\it {Earth-mass dark-matter haloes as the
  first structures in the early Universe}},  {\em Nature} {\bf 433} (2005)
  389--391, [\href{http://xxx.lanl.gov/abs/astro-ph/0501589}{{\tt
  astro-ph/0501589}}].

\bibitem{Frenk:2012ph}
C.~S. Frenk and S.~D.~M. White, {\it {Dark matter and cosmic structure}},  {\em
  Annalen Phys.} {\bf 524} (2012) 507--534,
  [\href{http://xxx.lanl.gov/abs/1210.0544}{{\tt arXiv:1210.0544}}].

\bibitem{Bromm:2009uk}
V.~Bromm, N.~Yoshida, L.~Hernquist, and C.~F. McKee, {\it {The formation of the
  first stars and galaxies}},  {\em Nature} {\bf 459} (2009) 49--54,
  [\href{http://xxx.lanl.gov/abs/0905.0929}{{\tt arXiv:0905.0929}}].

\bibitem{Umeda:2009yc}
H.~Umeda, N.~Yoshida, K.~Nomoto, S.~Tsuruta, M.~Sasaki, and T.~Ohkubo, {\it
  {Early Black Hole Formation by Accretion of Gas and Dark Matter}},  {\em
  JCAP} {\bf 0908} (2009) 024, [\href{http://xxx.lanl.gov/abs/0908.0573}{{\tt
  arXiv:0908.0573}}].

\bibitem{Colberg:2000zv}
{\bf VIRGO} Collaboration, J.~M. Colberg et~al., {\it {Clustering of galaxy
  clusters in CDM universes}},  {\em Mon. Not. Roy. Astron. Soc.} {\bf 319}
  (2000) 209, [\href{http://xxx.lanl.gov/abs/astro-ph/0005259}{{\tt
  astro-ph/0005259}}].

\bibitem{DiMatteo:2003zx}
T.~Di~Matteo, R.~A.~C. Croft, V.~Springel, and L.~Hernquist, {\it {Black hole
  growth and activity in a lambda CDM universe}},  {\em Astrophys. J.} {\bf
  593} (2003) 56--68, [\href{http://xxx.lanl.gov/abs/astro-ph/0301586}{{\tt
  astro-ph/0301586}}].

\bibitem{Maccio':2012uh}
A.~V. Maccio, O.~Ruchayskiy, A.~Boyarsky, and J.~C. Munoz-Cuartas, {\it {The
  inner structure of haloes in Cold+Warm dark matter models}},  {\em Mon. Not.
  Roy. Astron. Soc.} {\bf 428} (2013) 882--890,
  [\href{http://xxx.lanl.gov/abs/1202.2858}{{\tt arXiv:1202.2858}}].

\end{thebibliography}\endgroup

\end{document}